\documentclass[aps,prd,preprintnumbers,nofootinbib]{revtex4}
\usepackage[dvipdfmx]{graphicx}
\usepackage{bm,latexsym,amsmath,amsthm,amssymb,amsfonts,color}
\usepackage[normalem]{ulem}

\begin{document} 
\title{\uline{}The positive mass theorem in Kaluza-Klein picture}

\author{${}^{1,2}$Tetsuya Shiromizu and ${}^1$Diego Soligon}
\affiliation{${}^1$Department of Mathematics, Nagoya University, Nagoya 464-8602, Japan}
\affiliation{${}^2$Kobayashi-Maskawa Institute, Nagoya University, Nagoya 464-8602, Japan}

%%% To include the collaborator name... Please use the command "\collaborator"
%%% For example: \collaborator{ATLAS Collaboration}

\begin{abstract}%
We reconsider Schoen and Yau's proof of the positive mass theorem from the extra dimensional 
point of view, and we introduce a modified argument to prove the theorem in the Kaluza-Klein picture. 
We consider in this study an alternative condition to Jang's equation, which makes the argument more physically intuitive.
\end{abstract}

\maketitle

%
%======================================%
%<<<<<<<<<<<< SECTION I > >>>>>>>>>>>>>%
%======================================%
%
\section{Introduction} 

The positive mass theorem in general relativity asserts that for a nontrivial isolated physical system 
the total energy is nonnegative. It is fundamentally important as it guarantees the stability of space-time. 
Historically there have been two main ways to prove it. One of them, formulated by Witten \cite{Witten1981}, is inspired by supergravity  
and is easily accessible even by non-mathematicians. There has recently been some trial work based on this proof
on the construction of dark energy theories compatible with the positive mass theorem \cite{Nozawa2013, Nozawa2014, Tolley2014}, 
but these tend to be biased towards supergravity. Therefore, it would be interesting to address the same issue 
from another perspective, based on the other more mathematical proof of the positive mass theorem formulated by Schoen and Yau \cite{Schoen1979, Schoen1981},
which has no relation to supergravity. 
This approach however could cause some difficulty for the rather convoluted mathematical techniques required and the 
absence of an immediate physical interpretation. 

In this paper we propose a new method that follows the same argument as Schoen and Yau's proof \cite{Schoen1981},
but offers an easier interpretation which could have some applications in cosmology etc. In fact, Schoen and Yau's proof recalls
some characteristics of the Kaluza-Klein picture of space-time \cite{Kaluza1921,Klein1926}, an extra-dimensional theory first proposed 
as a possible candidate for a unified theory of gravity and electromagnetism. By reformulating the proof explicitly on 
a Kaluza-Klein space-time it is easier to obtain a physical interpretation.

The main idea of Ref.~\cite{Schoen1981} in extending the Riemannian positive mass theorem \cite{Schoen1979} to the 
general case is to consider a 
function $f$ on an initial data set  $(\Sigma, q_{ab}, K_{ab})$, where $q_{ab}$ and $K_{ab}$ are 
the induced metric and the second fundamental form of $\Sigma$, and take its graph $\bar \Sigma$ in $\Sigma\times \mathbb{R}$. 
Equipped with the product metric $dy^2+q$, the mean curvature of $\bar \Sigma$ in $\Sigma \times \mathbb{R}$ 
is supposed to be equal to $\bar q^{ab}K_{ab}$, a condition known as Jang's equation \cite{Jang1978}. 
It is noted that the induced metric on $\bar \Sigma$ can be deformed conformally to an asymptotically Euclidean metric with vanishing 
scalar curvature, so that the Riemannian positive mass theorem \cite{Schoen1979, Schoen2017} can be applied. 
%Notice also that $\bar\Sigma$
%can be identified with a linear slice of the Minkowski space-time when the ADM (Arnowitt-Deser-Misner) mass vanishes. 
We can then conclude by this result that the Arnowitt-Deser-Misner (ADM) mass of space-time is nonnegative.

The new approach we propose follows the same key steps, but instead of considering a function of an 
initial data set, we consider a function of space-time $M$, whose graph is a hypersurface in 
$M\times \mathbb{R}$, which can be regarded as a Kaluza-Klein space-time. In addition, the condition
imposed by Jang's equation is replaced by considering the existence of a marginally outer trapped surface (MOTS) in $M\times \mathbb{R}$. 
Analogies between solutions of Jang's equation and MOTS have already been considered \cite{Andersson2010}. 
The result is a new method of proving the positive mass theorem with an easy physical interpretation.

The rest of this paper is organized as follows. In Sec.~II we give a brief review of Schoen and Yau's 1981 
proof for non-experts. In Sec.~III we present a new way to prove the positive mass theorem in the 
Kaluza-Klein picture. Finally, we give a summary and a discussion of the result. 

%======================================%
%<<<<<<<<<<<< SECTION II  >>>>>>>>>>>>>%
%======================================%

\section{Brief review of the Positive Mass Theorem}

In this section, we will review Schoen and Yau's 1981 proof of the positive mass 
theorem \cite{Schoen1981} (See also Ref.~\cite{Eichmair2012}). This part will 
be helpful for non-experts. 
 
We consider a $n$-dimensional asymptotically flat initial data set for a space-time $(\Sigma,q_{ab},K_{ab})$, consisting of a $n$-dimensional
manifold $\Sigma$, a metric $q_{ab}$ and the second fundamental form $K_{ab}$, satisfying 
the constraint equations 
\begin{eqnarray}
 R-K_{ab}K^{ab}+K^2=2\rho \label{hami}
\end{eqnarray}
and
\begin{eqnarray}
D_aK^a_b-D_bK=J_b \,, \label{mom}
\end{eqnarray}
where $R$ is the scalar curvature of the metric $q_{ab}$, $\rho$ is the local mass density and $J_{b}$ is the local current density.
We assume that $\rho$ and $J_{b}$ obey the dominant energy condition
\begin{equation}
\rho\geq \left(J^b J_b\right)^{1/2} \,.
\end{equation}

We then form the $(n+1)$-dimensional product manifold $\Sigma\times \mathbb{R}$ with metric $\hat g$ defined by 
\begin{eqnarray}
\hat g=dy^2+q \,. 
\end{eqnarray}
In the above, we suppose that $q$ does not depend on $y$. 

Given a function $f$ on $\Sigma$, we consider a hypersurface $\bar \Sigma$ in $\Sigma\times \mathbb{R}$ 
which is the graph of the function $y=f(x^i)$, where $x^i$ is the coordinate on $\Sigma$ (See Fig. 1). Then the 
induced metric on $\bar \Sigma$ is 
\begin{eqnarray}
\bar q=\hat g|_{y=f(x)}=(q_{ij}+\partial_i f \partial_j f)dx^i dx^j=:\bar q_{ij}dx^i dx^j \,.
\end{eqnarray}

Hereafter we consider the ADM decomposition with respect to $\bar \Sigma$. 
The unit normal vector $\bar n_a$ to $\bar \Sigma$ in $\Sigma\times \mathbb{R}$ is 
\begin{eqnarray}
\bar n_a=\alpha \hat \nabla_a (y-f(x)) \,,
\end{eqnarray}
where $\alpha$ is the lapse function and $\hat \nabla_a$ is the covariant derivative with respect to $\hat g$.
In the current setup, we have 
\begin{eqnarray}
\alpha=(1+(Df)^2)^{-1/2} \,. 
\end{eqnarray}
The evolution equation for $\bar K$ along the $\bar n$-direction is given by
\begin{eqnarray}
\alpha^{-1}\bar D^2 \alpha=-\hat R_{ab}\bar n^a \bar n^b - \mbox \pounds_{\bar n} \bar K-\bar K_{ab} \bar K^{ab} \,,
\label{evo}
\end{eqnarray}
where $\hat R_{ab}$ is the Ricci tensor of $\hat g$, $\bar D_a$ is the covariant derivative with respect to the metric $\bar q$ 
and $\bar K_{ab}$ is the second fundamental form of $\bar \Sigma$. 

\begin{figure}
\begin{center}
\includegraphics[width=8cm]{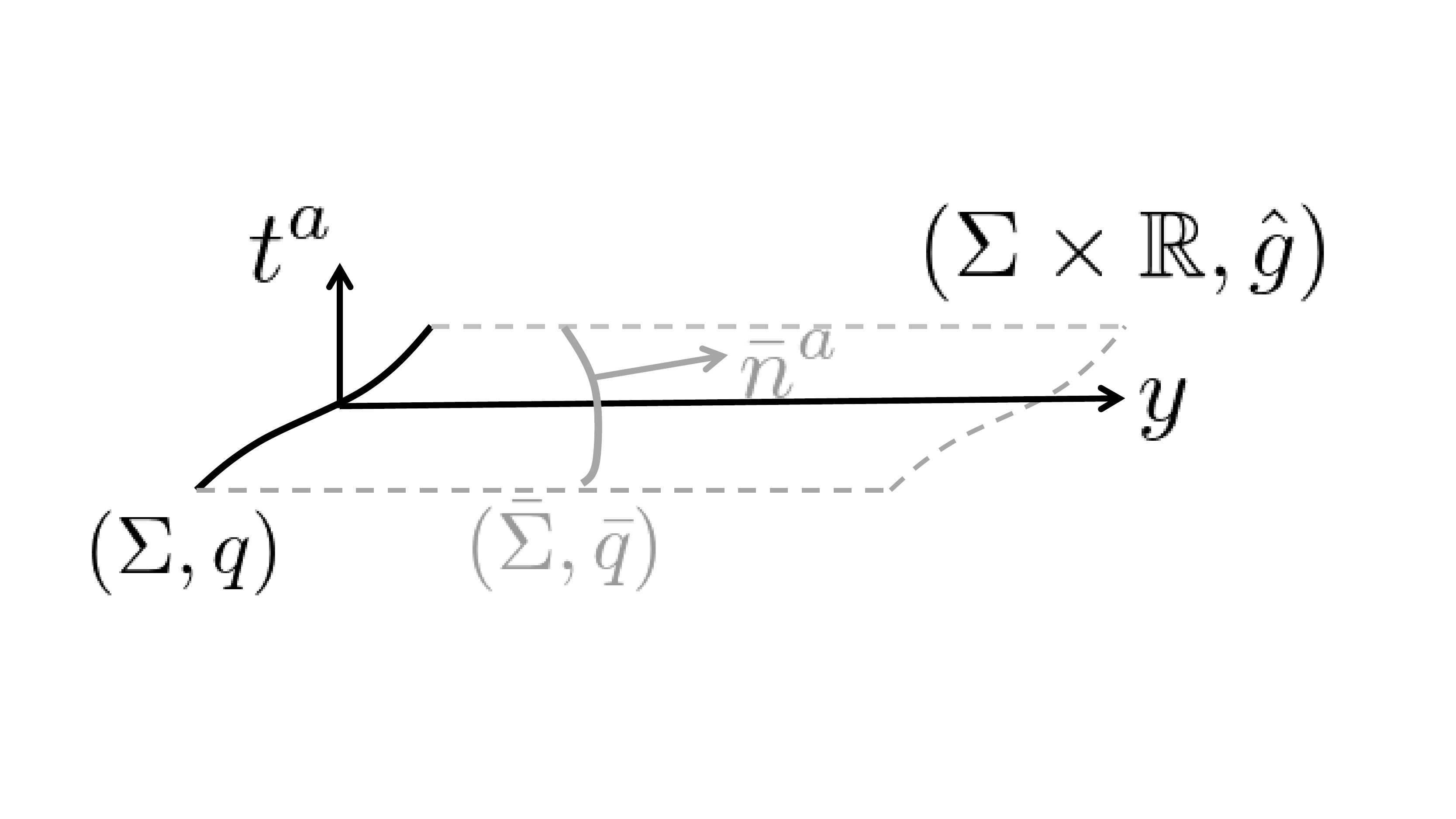}　
\caption{Setup in Schoen-Yau's 1981 proof}
\end{center}
\end{figure}

The double trace of the Gauss equation with respect to $\bar \Sigma$ gives us 
\begin{eqnarray}
\hat R-2\hat R_{ab}\bar n^a \bar n^b=\bar R-\bar K^2+\bar K_{ab} \bar K^{ab} \,,\label{doubleGausseq}
\end{eqnarray}
where $\bar R$ is the scalar curvature of $\bar \Sigma$. 
From the construction of $\hat g$, we see that $\hat R= R$. Using Eq.~(\ref{hami}), 
the equation above then becomes 
\begin{eqnarray}
2\hat R_{ab}\bar n^a \bar n^b=2\rho+K_{ab}K^{ab}-K^2-\bar K_{ab}\bar K^{ab}+\bar K^2-\bar R \,. \label{hami2}
\end{eqnarray}
Eqs.~(\ref{evo}) and (\ref{hami2}) imply 
\begin{eqnarray}
2\rho = \bar R-K_{ab}K^{ab}+K^2-\bar K^2-\bar K_{ab} \bar K^{ab}-2 \mbox \pounds_{\bar n} 
\bar K-2\alpha^{-1}\bar D^2 \alpha \,. \label{hami22}
\end{eqnarray}
Since 
\begin{eqnarray}
\hat \nabla_a (K^a_b-\delta^a_bK)=D_a(K^a_b-q^a_bK)  \label{codacci5tp4}
\end{eqnarray}
and direct calculation gives us 
\begin{eqnarray}
\bar n^b \hat \nabla_a (K^a_b-\delta^a_bK)=\bar n^a \hat \nabla_a K_{\bar q}-2\bar D^a \ln \alpha K_{ab} \bar n^b
-\bar D^a(\bar q_a^bK_{bc}\bar n^c)-\bar K K_{ab} \bar n^a \bar n^b+K_{ab} \bar K^{ab} \,,
\end{eqnarray}
Eq.~(\ref{mom}) implies 
\begin{eqnarray}
\mbox \pounds_{\bar n}K_{\bar q}-\bar D^a(\bar q_a^bK_{bc}\bar n^c)+J_a \bar n^a-\bar K K_{ab} \bar n^a \bar n^b+K_{ab} \bar K^{ab}
-2\alpha^{-1}K_{ab} \bar n^b \bar D^a \alpha=0 \,. \label{mom2}
\end{eqnarray}
Then, from Eqs.~(\ref{hami22}) and (\ref{mom2}) we derive the key equation 
\begin{eqnarray}
2(\rho-J_a \bar n^a)=-2\bar D^a X_a-2|X_a|^2_{\bar q}+\bar R+2\mbox \pounds_{\bar n}(K_{\bar q}-\bar K)
-|K_{ab}-\bar K_{ab}|^2_{\bar q}+K_{\bar q}^2-\bar K^2+2K_{ab}\bar n^a \bar n^b (K_{\bar q}-\bar K) \,, \label{syeq}
\end{eqnarray}
where $K_{\bar q}=\bar q^{ab}K_{ab}$ and 
\begin{eqnarray}
X_a:=\bar D_a \ln \alpha+\bar q_a^cK_{cb}\bar n^b \,.
\end{eqnarray}
$|\cdots|_{\bar q}$ denotes the trace with respect to $\bar q_{ab}$. 
Now, if one can impose 
\begin{eqnarray}
K_{\bar q}=\bar K \,, \label{jang}  
\end{eqnarray}
Eq.~(\ref{syeq}) becomes 
\begin{eqnarray}
2(\rho-J_a \bar n^a)=-2\bar D^a X_a-2|X_a|^2_{\bar q}+\bar R-|K_{ab}-\bar K_{ab}|^2_{\bar q} \,. \label{syeq2}
\end{eqnarray}

Eq. (\ref{jang}) can be written in term of $f$ as 
\begin{eqnarray}
\alpha \bar q^{ab}D_a D_bf=\bar q^{ab}K_{ab} \label{jang2}
\end{eqnarray}
and it is called Jang's equation.
It has been shown that a solution to Eq.~(\ref{jang2}) exists when there are no apparent horizons \cite{Schoen1981}. 
If there is an apparent horizon, a more careful treatment is needed. However, the essence of the 
proof does not depend on the existence of apparent horizons. Therefore, for simplicity, we focus on 
the case in which a solution to Jang's equation exists \footnote{If one is interested in the large scale structure of 
space-time, the assumption that there is no apparent horizon is reasonable.}. 

Let $\varphi$ be a function on $\bar \Sigma$. Let us multiply $\varphi^2$ to Eq.~(\ref{syeq2}) and integrate 
over $\bar \Sigma$. Using Eq.~(\ref{jang}), we have 
\begin{eqnarray}
\int_{\bar \Sigma} \Bigl[2(\rho-J_a \bar n^a) -\bar R\Bigr]\varphi^2 d\bar V
& = & \int_{\bar \Sigma}\Bigl[-2 \bar D^aX_a \varphi^2-2|X_a|^2_{\bar q}\varphi^2-|K_{ab}-\bar K_{ab}|^2_{\bar q}\varphi^2  
\Bigr]d\bar V \nonumber \\
& = & \int_{\bar \Sigma} \Bigl[ -2\left| \varphi X_a-\bar D_a \varphi\right|^2_{\bar q}-|K_{ab}-\bar K_{ab}|^2_{\bar q}+2(\bar D \varphi)^2 
\Bigr] d \bar V \nonumber \\
& \leq &  2 \int_{\bar \Sigma }(\bar D \varphi)^2 d \bar V \,. \label{intsyeq}
\end{eqnarray}
We suppose that $\varphi$ satisfies 
\begin{eqnarray}
\Bigl( \bar D^2 -\frac{n-2}{4(n-1)}\bar R \Bigr)\varphi=0 \label{3confscalar}
\end{eqnarray}
and has the following asymptotic behaviour at infinity
\begin{eqnarray}
\varphi = 1-C/r^{n-2}+O(1/r^{n-1}) \,.
\end{eqnarray}
Thus, by Eq.~(\ref{intsyeq}) and the dominant energy condition we have
\begin{eqnarray}
0 \leq 2 \int_{\bar \Sigma}(\rho-J_a \bar n^a)d \bar V 
\leq \int_{\bar \Sigma} \Bigl(\bar R+2(\bar D \varphi)^2 \Bigr)d \bar V
=\frac{4(n-1)}{n-2}\int_{\bar S_\infty}\varphi \bar D_a \varphi d \bar S^a
-\frac{2n}{n-2}\int_{\bar \Sigma}(\bar D \varphi)^2 d \bar V \,, 
\end{eqnarray}
hence 
\begin{eqnarray}
0 \leq \frac{2n}{n-2} \int_{\bar \Sigma}(\bar D \varphi)^2 d \bar V \leq  \frac{4(n-1)}{n-2}\int_{\bar S_\infty}\varphi \bar D_a \varphi d \bar S^a=64 \pi C \,, 
\end{eqnarray}
that is, $C \geq 0$. 

Take the conformal transformation $\tilde q_{ab}=\varphi^{4/(n-2)} \bar q_{ab}$. 
Eq.~(\ref{3confscalar}) shows us that the scalar curvature $\tilde R$ of $\tilde q$ vanishes. 
So, by the Riemannian positive mass theorem, the ADM mass $\tilde m$ is nonnegative \cite{Schoen1979, Schoen2017}. 
Since $m=\tilde m+2C$, we see that $m \geq 0$. Here we used the asymptotic behaviour of $q_{ij}$, that is, 
\begin{eqnarray}
q_{ij}=\Bigl(1+\frac{2}{n-2}\frac{m}{r^{n-2}} \Bigr)\delta_{ij}+O(1/r^{n-1}) \,.
\end{eqnarray}

We considered a $(n+1)$-dimensional product manifold in the proof, which can be interpreted as a spacelike slice of 
a Kaluza-Klein space-time, that is a $(n+2)$-dimensional space-time with $n+1$ space dimensions.
Jang's equation is a key point in the proof: its meaning is non-trivial at 
first glance. Here, Jang's equation can be written as 
\begin{eqnarray}
\bar q^{ab}(K_{ab}-\bar K_{ab})
=\frac{1}{2}\bar q^{ab}(\mbox \pounds_t q_{ab}-\mbox \pounds_{\bar n}\bar q_{ab})=0 \,. 
\label{janglie}
\end{eqnarray}
In consideration of the extra-dimensions construction, this expression suggests 
that the condition imposed by Jang's equation could be replaced by imposing the vanishing of 
the null expansion. The analogy between Jang's equation and the existence of a marginally outer 
trapped surface has been discussed extensively \cite{Andersson2010}. 

%======================================%
%<<<<<<<<<<<< SECTION III  >>>>>>>>>>>>>>%
%======================================%
%

\section{Proof of the Positive Energy Theorem in a Kaluza-Klein picture}

In this section we will slightly modify Schoen and Yau's proof of the positive mass 
theorem \cite{Schoen1981, Eichmair2012}. There are two main points in this new procedure. 
The first is that we shall consider the graph of a function on the full Lorentzian space-time instead of just on the 
Riemannian manifold corrisponding to the space dimensions. 
The second is the imposition of a condition alternative to Jang's equation. 
These are just slight modifications, but they allow for a better intuition of the physics behind the proof. 
Note that the notation in this section is independent of the previous one. 

Let $M$ be a $(n+1)$-dimensional Lorentzian space-time with metric $g_{\mu\nu}$, and  
consider a $(n+2)$-dimensional product manifold $\hat M=M\times \mathbb{R}$ equipped with the metric 
\begin{eqnarray}
\hat g=dy^2+g_{\mu\nu}dx^\mu dx^\nu \,,
\end{eqnarray}
where $g_{\mu\nu}$ does not depend on $y$. $(\hat M, \hat g_{ab})$ can be considered as a Kaluza-Klein 
space-time. Then, given a function $f$ on $M$ we can take a timelike hypersurface $\bar M$ given by the graph $y=f(x^\mu)$ (See Fig. 2). 
The metric induced on $\bar M$ is 
\begin{eqnarray}
\bar g_{\mu\nu}=g_{\mu\nu}+\partial_\mu  f \partial_\nu f \,. 
\end{eqnarray}
Introducing a change of coordinate defined by 
\begin{eqnarray}
\bar y:=y-f(x^\mu) \,,
\end{eqnarray}
the $(n+2)$-dimensional metric on $\hat M$ is written as 
\begin{eqnarray}
\hat g=d \bar y^2+2\partial_\mu f d\bar y dx^\mu+\bar g_{\mu\nu}dx^\mu dx^\nu \,. 
\end{eqnarray}
The unit normal vector to $\bar M$ in $\hat M$ is given by 
\begin{eqnarray}
\bar n_a=\bar \alpha \hat\nabla_a \bar y \,,
\end{eqnarray}
where $\bar \alpha=(1+ g^{\mu\nu}\partial_\mu f \partial_\nu f )^{-1/2}$ and $\hat \nabla_a$ is the 
covariant derivative with respect to $\hat g_{ab}$. Here we suppose $g^{\mu\nu}\partial_\mu f \partial_\nu f>0$ because 
$\bar M$ is a timelike hypersurface. 

Consider a spacelike hypersurface $(\bar \Sigma, \bar q_{ab}) \subset (\bar M, \bar g_{ab})$ given by 
the intersection of $\bar M$ and the spacelike hypersurface $\hat \Sigma$ with timelike unit normal vector $\bar t^a$ 
in $\hat M$, that is $\bar \Sigma=\bar M \cap \hat \Sigma$ (See Fig. 2). Then, the metric is decomposed as 
\begin{eqnarray}
\hat g_{ab}=\bar g_{ab}+\bar n_a \bar n_b
=\bar h_{ab}-\bar t_a \bar t_b
=\bar q_{ab}-\bar t_a \bar t_b+\bar n_a \bar n_b \,, \label{metricdecomp}
\end{eqnarray}
where $\bar g_{ab}$, $\bar h_{ab}$ and $\bar q_{ab}$ are the induced metrics on $\bar M$, $\hat \Sigma$ and 
$\bar \Sigma$ respectively.  
\begin{figure}
\begin{center}
\includegraphics[width=8cm]{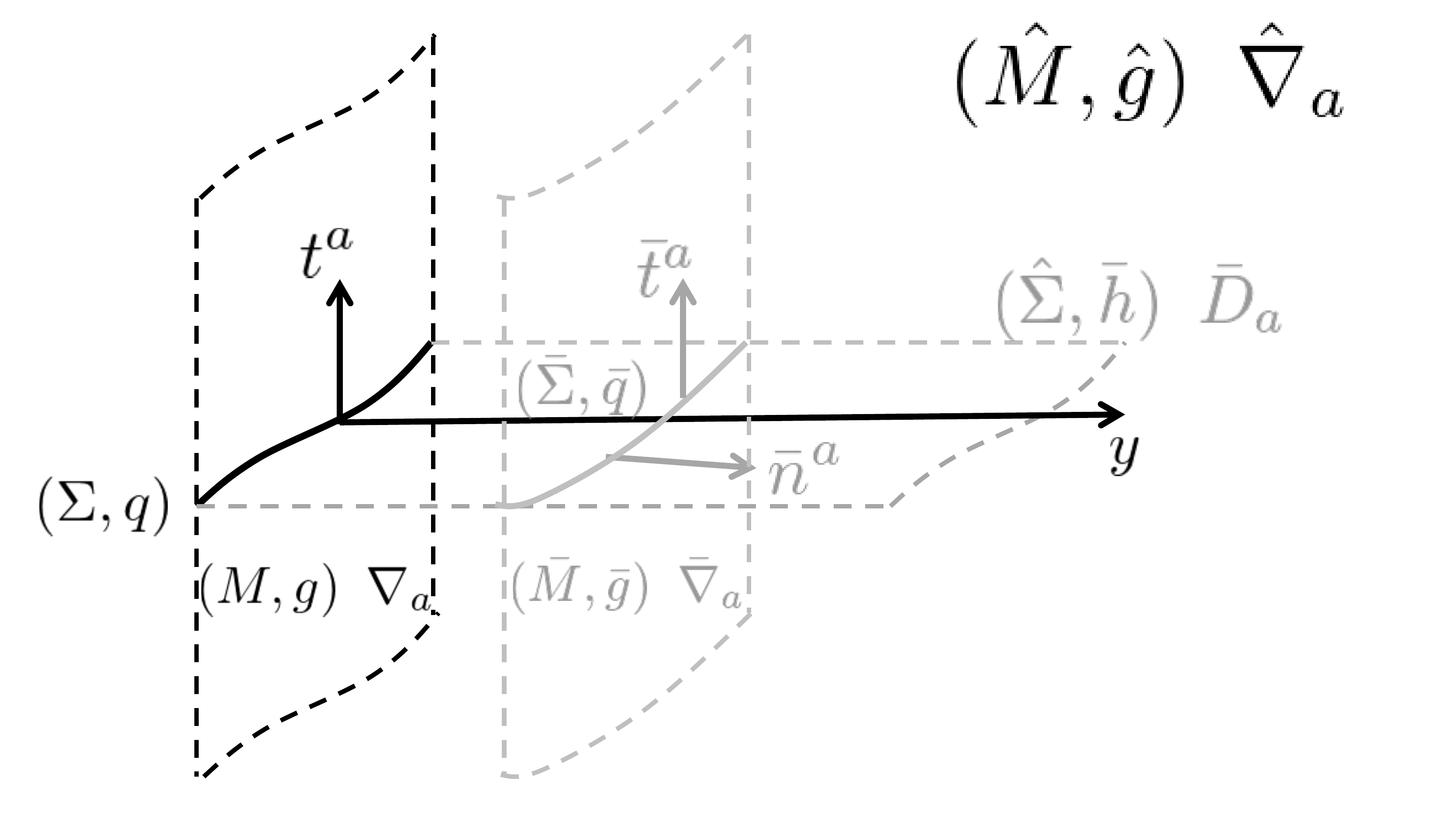}　
\caption{Setup in our consideration}
\end{center}
\end{figure}
We denote the covariant derivatives $\bar  D_a$ and $\bar {\cal D}_a$ with 
respect to $\bar h_{ab}$ and $\bar q_{ab}$ respectively. 
Then, we have following key equation \cite{Andersson2010,Galloway2005}
\begin{eqnarray}
\bar n^a \bar D_a \bar \theta_-=\frac{1}{2}\bar \theta_{-ab} \bar \theta_-^{~~ab}-\frac{1}{2} \bar \theta_-^2
+(k_{(\bar n)}- \kappa_{(\bar t)}) \bar \theta_-+\bar {\cal D}_a \bar \tau^a
+\bar \tau_a \bar \tau^a-\frac{1}{2}R_{\bar q}+\hat G_{ab}\bar \ell^a \bar t^b \,, \label{cross}
\end{eqnarray}
where 
\begin{eqnarray}
\bar \theta_- := \bar q^{ab} \hat\nabla_a (\bar t_b-\bar n_b) \,,
\end{eqnarray}
\begin{eqnarray}
\bar \theta_{-ab} := \bar q_a^{ c} \bar q_b^{ d} \hat \nabla_c (\bar t_b-  \bar n_b) \,,
\end{eqnarray}
\begin{eqnarray}
k_{(\bar n)}:=\bar q^{ab} \bar D_a \bar n_b \,,
\end{eqnarray}
\begin{eqnarray}
\kappa_{(\bar t)}:=\bar n^a \bar n^b \hat \nabla_a \bar t_b \,,
\end{eqnarray}
\begin{eqnarray}
\bar \tau_a:=\bar q_a^{ b} \hat\nabla_b \bar t_c \bar n^c+\bar {\cal D}_a \ln \bar \alpha \,,
\end{eqnarray}
\begin{equation}
\bar \ell^a:=\bar t^a-\bar n^a
\end{equation}
with a null vector $\bar\ell^a$, and where $R_{\bar q}$ is the scalar curvature of $\bar q_{ab}$ and  $\hat G_{ab}$ is the Einstein tensor 
for $\hat g_{ab}$.  

In the current setup, it is easy to see that
\begin{eqnarray}
\hat G_{ab}\bar \ell^a \bar t^b=G_{ab}\bar \ell^a \bar t^b \,, \label{2einsteintensor}
\end{eqnarray}
where $G_{ab}$ is the Einstein tensor for $g_{ab}$. Since $\bar \ell^a$ is a null vector 
in $\hat M$, Eq. (\ref{metricdecomp}) tells us 
\begin{eqnarray}
g_{ab}\bar \ell^a \bar \ell^b =-(n_a \bar \ell^a)^2 \,,
\end{eqnarray}
so that $g^a_{ b} \bar \ell^b$ is a timelike vector in $M$. In a similar way, $g^a_{ b} \bar t^b$ 
is also a timelike vector. Imposing the dominant energy condition for the Einstein equation on $(M,g)$, 
Eq.~(\ref{2einsteintensor}) implies 
\begin{eqnarray}
\hat G_{ab}\bar \ell^a \bar t^b \geq 0 \,.
\end{eqnarray} 

Let us now introduce the condition that $\bar \Sigma$ is required to satisfy, which corresponds to Jang's equation in the original proof: 
\begin{eqnarray}
\bar \theta_-=\bar q^{ab} \hat\nabla_a \bar t_b- \bar q^{ab} \hat\nabla_a \bar n_b=0 \,. \label{neweq}
\end{eqnarray}
In Jang's equation as presented by \cite{Schoen1981}, 
the first term is $\bar q^{ab} \hat \nabla_a  t_b$, where 
$\hat \nabla_a  t_b$ is treated as the pullback from $\Sigma$ to $\bar \Sigma$ (See Appendix A for 
more details on the distinction between the original proof and the one presented here). Note that Eq.~(\ref{neweq}) is 
satisfied by a marginally outer trapped surface \cite{Andersson2010}. 
Note also that $\bar \Sigma$ is a non-compact, asymptotically flat surface. 
In terms of $\bar y$, the equation becomes
\begin{eqnarray}
\bar \alpha \bar q^{ab} \hat\nabla_a \hat\nabla_b \bar y=\bar q^{ab} \hat\nabla_a 
\bar t_b \,. \label{nulljang}
\end{eqnarray}
This is an elliptic equation for $\bar y$. The existence of a solution to 
Jang's equation was proved by Schoen and Yau \cite{Schoen1981, Eichmair2012}
if there is no apparent horizon in $\bar \Sigma$ 
and it was a fundamental part of their result. As for the existence of a solution to Eq.~(\ref{nulljang}),
the question will be left open in this paper and we will work under the assumption that it exists. 

Once we impose Eq. (\ref{nulljang}), we see that on $\bar \Sigma$
\begin{eqnarray}
\bar n^a \bar D_a \bar \theta_-=0 \,.
\end{eqnarray}
Since $\bar \Sigma$ is not compact, this does not imply that 
$\bar \Sigma$ is an apparent horizon, however one may regard $\bar \Sigma$ as having a similar 
role to an apparent horizon from a technical point of view. 

Let $\varphi$ be a function over $\bar \Sigma$, satisfying 
\begin{eqnarray}
\Bigl( \bar {\cal D}^2-\frac{n-2}{4(n-1)}R_{\bar q} \Bigr) \varphi=0 \,,\label{scalar}
\end{eqnarray}
with asymptotic behaviour 
\begin{eqnarray}
\varphi=1-C/r^{n-2}+O(1/r^{n-1}) \,.
\end{eqnarray}
As in the previous section, we multiply Eq.~(\ref{cross}) by $\varphi^2$ and integrate over $\bar \Sigma$, and by the dominant energy condition 
and Eq.~(\ref{neweq}), 
\begin{eqnarray}
\int_{\bar \Sigma} \varphi^2 
\Bigl(\frac{1}{2}R_{\bar q}-\bar \tau_a \bar \tau^a -\bar {\cal D}_a \bar \tau^a \Bigr)
d \bar V \geq 0 \,.
\end{eqnarray}
Using Eq.~(\ref{scalar}), by the Gauss theorem we have
\begin{eqnarray}
\int_{\bar S_\infty}\Bigl( \frac{2(n-1)}{n-2}\varphi \bar {\cal D}_a \varphi-\varphi^2 \bar \tau_a \Bigr)d \bar S^a 
\geq \int_{\bar \Sigma} \Bigl[\frac{n}{n-2}(\bar {\cal D} \varphi)^2
+(\varphi \bar \tau_a-\bar {\cal D}_a \varphi )^2 \Bigr] d \bar V \geq 0 \,. \label{intcross}
\end{eqnarray}
It is easy to see that $\bar \tau_a =O(1/r^{2n-3})$ does not contribute to the surface integral in the 
right-hand side, therefore 
\begin{eqnarray}
C \geq 0 \,. \label{cineq}
\end{eqnarray}

Now perform a conformal transformation given by 
\begin{eqnarray}
\tilde  q=\varphi^{4/(n-2)} \bar q \,. 
\end{eqnarray}
Asymptotically at infinity 
\begin{eqnarray}
\bar q_{ij}=\Bigl(1+ \frac{2}{n-2}\frac{\bar m}{r^{n-2}}\Bigr)\delta_{ij}+O(1/r^{n-1})
\end{eqnarray}
and
\begin{eqnarray}
\tilde q_{ij}=\Bigl(1+\frac{2}{n-2}\frac{\tilde m}{r^{n-2}}  \Bigr)\delta_{ij}+O(1/r^{n-1}) \,,
\end{eqnarray}
where $\bar m$ and $\tilde m$ are the ADM masses for $\bar g$ and $\tilde g$ respectively. 
The conformal transformation tells us 
\begin{eqnarray}
\bar m= \tilde m+2C \label{massrelation}
\end{eqnarray}
and 
\begin{eqnarray}
\tilde R = 0 \,.
\end{eqnarray}
Therefore we can apply the Riemannian positive mass theorem \cite{Schoen1979, Schoen2017} to $(\tilde \Sigma, \tilde q)$ 
so that 
\begin{eqnarray}
\tilde m \geq 0 \,. 
\end{eqnarray}
Hence, by Eq.~(\ref{massrelation}) 
\begin{eqnarray}
\bar m \geq 0 \,. 
\end{eqnarray}
Since $\bar q_{ij}=q_{ij}+\partial_i f \partial_j f$ with $\partial_i  f =O(1/r^{n-1})$, 
$m=\bar m$. Thus, the ADM mass of spacetime is nonnegative. 

\section{Summary and discussion}

In this paper we propose an alternative proof of the positive mass theorem in a Kaluza-Klein picture. Instead of considering
the graph of a function on a Riemannian manifold, we consider the graph of a function on the full Lorentzian space-time, which can be considered 
as a hypersurface on a Kaluza-Klein space-time. Jang's equation is replaced by a condition directly related to the existence of a marginally trapped outer surface.
Compared to the original proof, this proof provides a more direct physical intuition, paving the way for
future consideration in the field of cosmology for example.

An open question that remains is the existence of a solution to Eq.~(\ref{nulljang}), which was assumed in this paper.
Another issue is the case in which there is an apparent horizon: that would need a careful treatment. Considering the similarity to Jang's equation
we would expect to be able to use a similar argument in this case.

\begin{acknowledgments}
We would like to thank Tatsuya Morino for his presentation of Schoen and Yau's proof of the positive mass theorem. 
We would also like to thank Prof.~Sumio Yamada for the useful discussion on Jang's equation. 
T. S. is supported by Grant-Aid for Scientific Research from Ministry of Education, 
Science, Sports and Culture of Japan (No. 16K05344, 17H01091). This work is supported 
in part by JSPS Bilateral Joint Research Projects (JSPS-NRF collaboration) ``String 
Axion Cosmology". 
\end{acknowledgments}

\appendix

\section{ADM-decomposition}

To see how Jang's equation and the one proposed in this paper are explicitly distinct, consider the ADM decomposition of
the metrics $g$ of $M$ and $\bar g$ of $\bar M$:  
\begin{eqnarray}
g=g_{\mu\nu}dx^\mu dx^\nu =- N^2 dt^2+ q_{ij}(dx^i+ N^idt)(dx^j+ N^j dt) \,,
\end{eqnarray}
\begin{eqnarray}
\bar g=\bar g_{\mu\nu}dx^\mu dx^\nu =(g_{\mu\nu}+\partial_\mu f \partial_\nu f)dx^\mu dx^\nu
=-\bar N^2 dt^2+\bar q_{ij}(dx^i+\bar N^idt)(dx^j+\bar N^j dt) \,.
\end{eqnarray}
The timelike unit normal vectors to $\Sigma$ and $\bar \Sigma$ are written as
\begin{eqnarray}
t^a=N^{-1}[(\partial_t)^a-N^i (\partial_i)^a]
\end{eqnarray}
and
\begin{eqnarray}
\bar t^a=\bar N^{-1}[(\partial_t)^a-\bar N^i (\partial_i)^a] \,.
\end{eqnarray}
A direct comparison tells us 
\begin{eqnarray}
-\bar N^2+\bar q_{ij}\bar N^i \bar N^j=-N^2+q_{ij}N^i N^j+\dot f^2 \,,
\end{eqnarray}
\begin{eqnarray}
\bar q_{ij}\bar N^j=q_{ij}N^j+\dot f f_i
\end{eqnarray}
and
\begin{eqnarray}
\bar q_{ij}=q_{ij}+f_i f_j \,,
\end{eqnarray}
where $\dot f=\partial_t f$ and $f_i=\partial_i f$. 

Noting that the inverse of $\bar g_{\mu\nu}$ is given by
\begin{eqnarray}
\bar g^{\mu\nu}=g^{\mu\nu}-\bar \alpha^2 \nabla^\mu f \nabla^\nu f \,,
\end{eqnarray}
the relation between $N^i$ and $\bar N^i$ becomes
\begin{eqnarray}
\bar N^i =N^i -\bar \alpha^2 {\cal D}^i f f_N+\dot f (1-\bar \alpha^2 f_N^2) \,, \label{nibarni}
\end{eqnarray}
where $f_N=N^i f_i$. For $N$ and $\bar N$, we have 
\begin{eqnarray}
\bar N^2=N^2+f_N^2+\bar \alpha^2 f_N^2[1+({\cal D} f)^2][\bar \alpha^2 ({\cal D}f)^2-2]
+\dot f^2[1+({\cal D} f)^2][({\cal D} f)^2+\bar \alpha^2f_N^4 (\bar \alpha^2 ({\cal D}f)^2-2) ] \,. \label{nbarn}
\end{eqnarray}

When 
\begin{eqnarray}
\dot f=0 \,,
\end{eqnarray}
since
\begin{eqnarray}
\bar \alpha^2=(1+({\cal D} f)^2)^{-1/2} \,,
\end{eqnarray}
Eqs.~(\ref{nibarni}) and (\ref{nbarn}) are simplified to 
\begin{eqnarray}
\bar N^i=N^i-\bar \alpha^2 {\cal D}^i f f_N
\end{eqnarray}
and
\begin{eqnarray}
\bar N^2=N^2-\bar \alpha^2f_N^2 \,.
\end{eqnarray}

Since $t^a$ and $\bar t^a$ are different quantities, 
$\bar q^{ab} \hat \nabla_a \bar t_b$ and $\bar q^{ab} \hat \nabla_c t_b$ which appear 
in the paper are also different quantities.

\end{document}